\documentclass[aps,superscriptaddress,pre,reprint,amsmath,amssymbols]{revtex4-1}
\usepackage{graphicx}
\usepackage{array}
\usepackage{booktabs}
\usepackage[colorlinks, linkcolor=blue]{hyperref}
\usepackage{bm}
\usepackage{color}
\usepackage{subfigure}
\usepackage{soul}
\usepackage{amsmath}
\usepackage{amsfonts}
\usepackage{amssymb}
\usepackage{lipsum}
\usepackage{calrsfs}
\DeclareMathAlphabet{\pazocal}{OMS}{zplm}{m}{n}

\newcommand{\DpRP}{\ensuremath{\delta_\textsc{rp}}}
\newcommand{\DpB}{\ensuremath{\delta_\textsc{j}}}
\newcommand{\kB}{\ensuremath{k_\textsc{b}}}
\newcommand{\kN}{\ensuremath{k_\textsc{n}}}
\newcommand{\pB}{\ensuremath{p_\textsc{b}}}
\newcommand{\pN}{\ensuremath{p_\textsc{n}}}
\newcommand{\XL}{\ensuremath{\chi_\textsc{l}}}
\newcommand{\XT}{\ensuremath{\chi_\textsc{t}}}

\definecolor{darkBlue}{rgb}{0,0,0.6}
\definecolor{darkRed}{rgb}{0.5,0,0}
\definecolor{darkGreen}{rgb}{0,0.5,0}

\begin{document}
\title{Universal scaling for disordered viscoelastic matter II: \\ Collapses, global behavior and spatio-temporal properties}
\author{Danilo B. Liarte}
\email{liarte@cornell.edu}
\affiliation{Institute of Physics, University of S\~ao Paulo, S\~ao Paulo, SP, Brazil}
\affiliation{ICTP South American Institute for Fundamental Research, S\~ao Paulo, SP, Brazil}
\affiliation{Instituto de F\'isica Te\'orica, Universidade Estadual Paulista, S\~ao Paulo, SP, Brazil}
\affiliation{Department of Physics, Cornell University, Ithaca, NY 14853, USA}
\author{Stephen J. Thornton}
\affiliation{Department of Physics, Cornell University, Ithaca, NY 14853, USA}
\author{Eric Schwen}
\affiliation{Department of Physics, Cornell University, Ithaca, NY 14853, USA}
\author{Itai Cohen}
\affiliation{Department of Physics, Cornell University, Ithaca, NY 14853, USA}
\author{Debanjan Chowdhury}
\affiliation{Department of Physics, Cornell University, Ithaca, NY 14853, USA}
\author{James P. Sethna}
\affiliation{Department of Physics, Cornell University, Ithaca, NY 14853, USA}
\date{\today}
	
\begin{abstract}
Disordered viscoelastic materials are ubiquitous and exhibit fascinating invariant scaling properties.
In a companion article~\cite{LiarteSet2021}, we have presented comprehensive new results for the critical behavior of the dynamic susceptibility of disordered elastic systems near the onset of rigidity.
Here we provide additional details of the derivation of the singular scaling forms of the longitudinal response near both jamming and rigidity percolation.
We then discuss global aspects associated with these forms, and make scaling collapse plots for both undamped and overdamped dynamics in both the rigid and floppy phases.
We also derive critical exponents, invariant scaling combinations and analytical formulas for universal scaling functions of several quantities such as transverse and density responses, elastic moduli, viscosities, and correlation functions.
Finally, we discuss tentative experimental protocols to measure these behaviors in colloidal suspensions.
\end{abstract}

\maketitle

\section{Introduction}
\label{sec:Introduction}

Disordered elastic systems encompass a wide range of materials, from amorphous solids~\cite{Wyart2005} and network glasses~\cite{Thorpe1983} to biopolymer fiber networks~\cite{BroederszMac2011}, articular cartilage~\cite{JacksonCoh2022}, confluent cell tissues~\cite{BiMan2015,BiMan2016} and even machine learning~\cite{BahriGan2020}.
Their theoretical development has not only led to a much deeper understanding of traditionally difficult problems as the glass transition; it has also pushed the boundaries of science to incorporate new frameworks such as topological mechanics~\cite{BertoldiHec2017,MaoLub2018}, non-reciprocal phase transitions~\cite{FruchartVit2020} and novel mechanical metamaterials~\cite{GoodrichNag2015,RocksNag2017,ChenMah2019,LiarteLub2020}.
In a companion article~\cite{LiarteSet2021}, we have employed a systematic analysis of the invariant scaling of the dynamic susceptibility to determine the universal critical behavior of several classes of disordered viscoelastic materials near the onset of rigidity.
Here we present a derivation of the theoretical results shown in~\cite{LiarteSet2021}, and discuss additional details for scaling collapses, the global behavior of universal scaling functions, and general scaling forms for diverse spatio-temporal properties such as moduli, viscosities and correlation functions.

\emph{Jamming}~\cite{LiuNag2010} and \emph{Rigidity Percolation} (RP)~\cite{Thorpe1983} provide two of the most suitable approaches to characterize the fascinating scaling behavior that is exhibited by several classes of disordered viscoelastic materials near the onset of rigidity~\cite{SethnaZap2017}.
Both are often modeled by elastic networks close to Maxwell's threshold for mechanical stability~\cite{Maxwell1864}, and represent transitions from a rigid phase to a floppy one when the average coordination number becomes smaller than the isostatic value.
RP is usually described in terms of networks in which bonds between sites are randomly removed, and is characterized by a second-order transition for all elastic moduli~\cite{FengGar1985,LiarteLub2016}.
In turn, jamming is usually described in terms of disordered arrangements of spheres that exhibit an unusual critical behavior, with a first-order transition for the bulk modulus and a second-order transition for the shear modulus.

Whereas effective-medium theories have been routinely employed in the derivation of the critical behavior associated with RP~\cite{FengGar1985,MaoLub2011,LiarteLub2016}, a genuine finite-dimensional effective-medium theory for jamming had remained elusive until recently~\cite{LiarteLub2019}.
Incidentally, a phenomenological theory providing a synthesis of available numerical work on the universal critical scaling of jamming has also been proposed recently~\cite{GoodrichSet2016}.
Here we show how to combine these two developments~\cite{LiarteLub2019,GoodrichSet2016} to derive the scaling behavior of a large class of disordered elastic materials near jamming and rigidity percolation.
Our results are based on a scaling Ansatz for the longitudinal response that is akin to the one considered in Ref.~\cite{GoodrichSet2016}.
They go beyond the results of~\cite{GoodrichSet2016} by incorporating rigidity percolation in addition to jamming, as well as wavelength and frequency dependencies in addition to static results.
We use the effective-medium theory of Ref.~\cite{LiarteLub2019} to both validate our scaling forms and to extract analytical formulas for the universal scaling functions.
Our results go beyond the results of~\cite{LiarteLub2019} by incorporating the analysis of a wide variety of physical quantities, particularly spatio-temporal properties such as dynamic response and correlation functions.

The remainder of this article is organized as follows.
In Sec.~\ref{sec:Theory}, we present a brief review of key results of Ref.~\cite{LiarteLub2019} (Sec.~\ref{subsec:Review}), and a derivation of the critical exponents and universal scaling functions for the longitudinal response (Sec.~\ref{subsec:Derivation}).
To validate the theory of Sec.~\ref{subsec:Derivation}, we present in Sec.~\ref{sec:Global} scaling collapse plots near both jamming and rigidity percolation, for undamped and overdamped dynamics, in the rigid and floppy phases, and discuss the global behavior of the longitudinal response.
We then use results from Sec.~\ref{subsec:Derivation} to derive the universal scaling behavior of several additional quantities in Sec.~\ref{sec:Quantities} --- the transverse dynamic response, moduli and viscosities, density response and correlation functions. 
Finally, we end with an outlook in Sec.~\ref{sec:Summary}.

\section{Theory}
\label{sec:Theory}

Here we present a brief review of key results from Ref.~\cite{LiarteLub2019}, which we use in the derivation of critical exponents and universal scaling functions for the longitudinal response (Sec.~\ref{subsec:Derivation}) and other quantities (Sec.~\ref{sec:Quantities}).

\subsection{Effective-medium theory for jamming and rigidity percolation}
\label{subsec:Review}

\begin{figure}[!ht]
	\centering
	\includegraphics[width=0.8\linewidth]{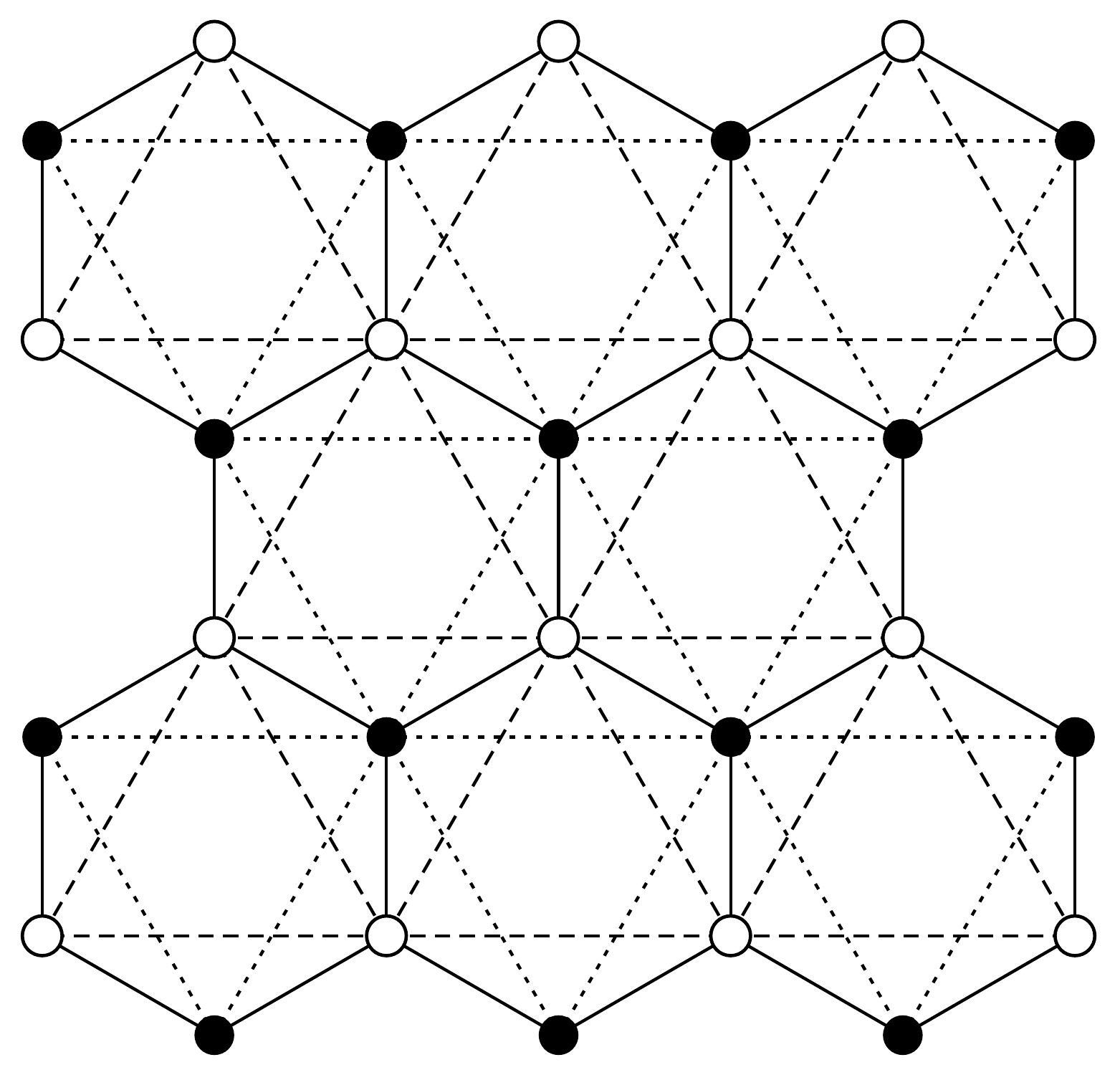}
	\caption{Illustration of the honeycomb-triangular lattice model. In simulations, bonds of the honeycomb lattice (solid lines) and the two triangular lattices (dashed and dotted lines) are populated with probability $\pB$ and $\pN$, respectively.
	In the effective medium theory, the fully-occupied honeycomb and triangular	sub-lattices have frequency-dependent effective spring constants $\kB(\omega)$ and $\kN(\omega)$, respectively, satisfying a set of self-consistent equations.}
	\label{fig:HTL}
\end{figure}

We use the honeycomb-triangular lattice (HTL) model~\cite{LiarteLub2019} to describe both static and dynamical properties of jamming and rigidity percolation (RP) near the threshold of mechanical stability.
The HTL model combines suitable properties of two periodic lattices: The honeycomb lattice (solid lines in Fig.~\ref{fig:HTL}) with finite bulk modulus $B>0$ and zero shear modulus $G=0$, and two triangular lattices (dashed and dotted lines) with finite $B,G>0$.
In simulations, bonds (harmonic elastic interactions) of the honeycomb lattice and the triangular lattices have unit spring constant and are populated~%
\footnote{This is a special case of the more general model in which the bond occupation probability can be different for the two triangular lattices.}
with probability $\pB$ and $\pN$, respectively.
In the effective-medium theory, the honeycomb and triangular lattices are fully populated with bonds with complex frequency-dependent effective spring constant $\kB(\omega)$ and $\kN(\omega)$, respectively.

We use the Coherent Potential Approximation~\cite{ElliottLea1974,FengGar1985,MaoLub2011} (CPA) to derive a set of self-consistent equations for $\kB(\omega)$ and $\kN(\omega)$, and then describe elastic and phonon properties of the HTL model near jamming and RP.
The derivation of the CPA self-consistent equations is standard, and we do not include it here.
Here the randomly-diluted lattice is modeled by homogeneous lattices with effective spring constants satisfying the set of equations~\cite{LiarteLub2019}:
\begin{equation}
	k_\alpha =
		\frac{p_\alpha - h_\alpha}{1-h_\alpha},
	\label{eq:cpa}
\end{equation}
where $p_\alpha$ and $k_\alpha$ are the occupancy probability of each bond and the effective spring constant for sub-lattice $\alpha \in \{B,N\}$, respectively.
The functions $h_\alpha = h_\alpha (\pB, \pN, \omega)$ are defined by
\begin{equation}
	h_\alpha =
		\frac{1}{\tilde{z}_\alpha N_c} \sum_{\bm{q}} \text{Tr} \left[
			D_\alpha (\bm{q}) \cdot \mathcal{G} (\bm{q}, \omega) \right],
\end{equation}
where $N_c$ is the total number of cells, $\omega$ is the frequency, $q$ is the wavevector, and $D_\alpha (\bm{q})$ and $\tilde{z}_\alpha$ are the dynamical matrix and the number of bonds per unit cell for sub-lattice $\alpha$, respectively.
The trace is taken over an $m D$-dimensional space, where $m$ is the number of sites per unit cell and $D$ is the spatial dimension.
The Green's function $\mathcal{G}$ is defined by
\begin{equation}
	\mathcal{G} (\bm{q}, \omega) =
		\left[\sum_{\alpha}D_\alpha (\bm{q})
			- \omega^2 \mathbb{I} \right]^{-1},
\end{equation}
where $\mathbb{I}$ is an identity matrix.
Note that $\sum_\alpha D_\alpha$ depends on all effective spring constants $k_\alpha$, so that Eq.~\ref{eq:cpa} self-consistently determines the values of all $k_\alpha$ for given $\pB$, $\pN$ and $\omega$.

\begin{figure}[!ht]
	\centering
	\includegraphics[width=0.8\linewidth]{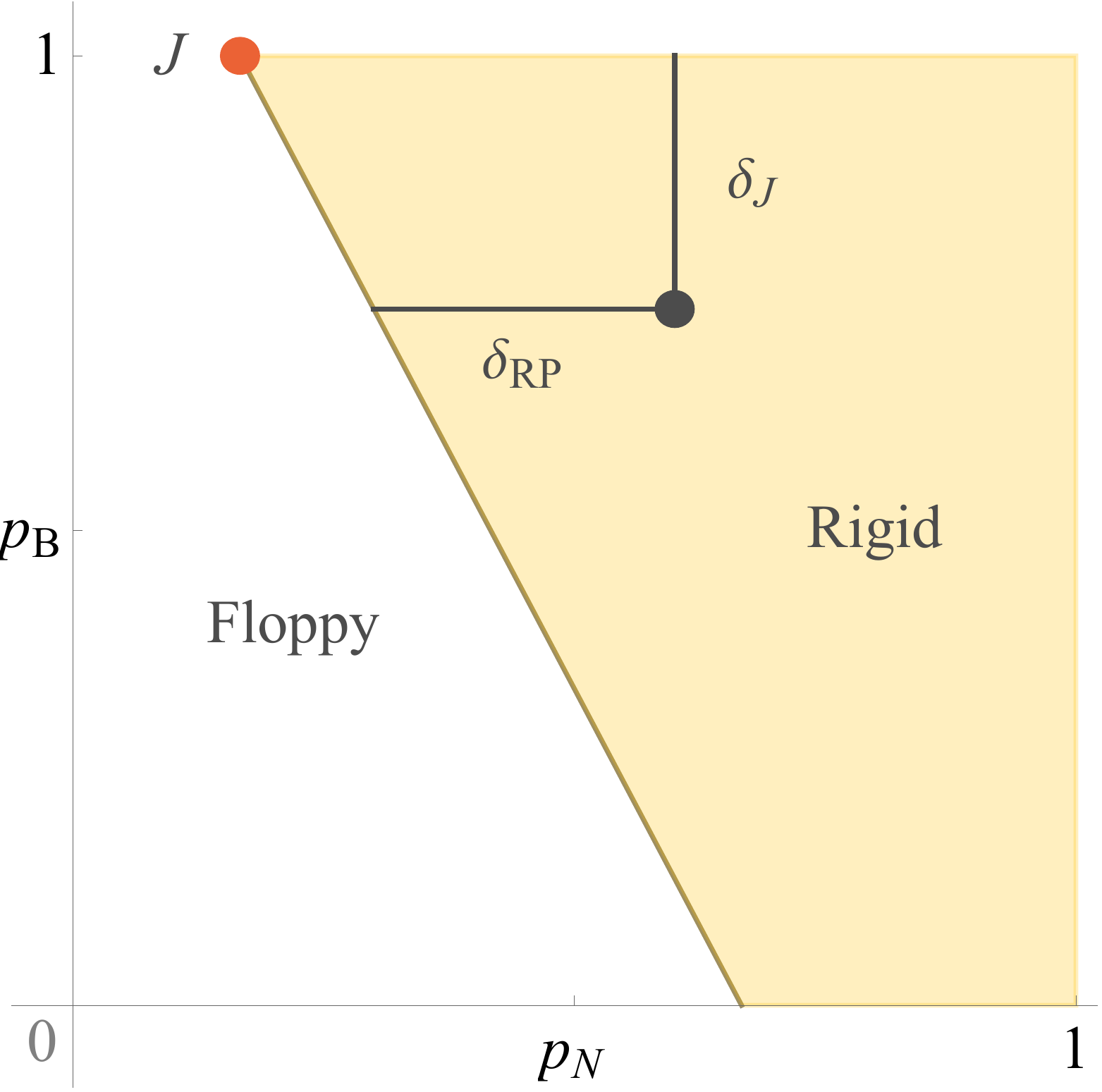}
	\caption{Phase diagram of the HTL model in terms of bond occupation probabilities for the honeycomb ($\pB$) and triangular ($\pN$) lattices.
	The gray line is an RP line that terminates at a multicritical jamming point $J$ (red disk).
	The diagram also shows the pair of scaling variables $\DpB$ and $\DpRP$.}
	\label{fig:PhaseDiagram}
\end{figure}

Elastic moduli can be expressed in terms of the effective springs constants by taking the long-wavelength limit of the dynamical matrix while ensuring that internal degrees of freedom are relaxed before the limit of small wavevector is taken.
The HTL has isotropic elasticity, with bulk and shear moduli given by
\begin{equation}
	B =
		\frac{3}{4} \, \kB + \frac{9}{2} \, \kN, \quad
	G =
		\frac{9}{4}\, \kN,
\end{equation}
respectively.
These definitions along with solutions of the CPA equations allow us to draw the zero-frequency phase diagram shown in Fig. \ref{fig:PhaseDiagram}.

Using a perturbation analysis, one can write the asymptotic equations for the low-frequency behavior of $\kB$ and $\kN$ near the jamming point $J$~\cite{LiarteLub2019}:
\begin{align}
    & \kB
        \approx \frac{\kN}{\kN+\DpB / b_1},
    \label{eq:kb_asymptotic} \\
    & \kN
        \approx b_2 \, |\DpRP| \left( \sqrt{1-c \, \frac{\tilde{\omega} (\omega)}{|\DpRP|^2}} \pm 1 \right),
    \label{eq:kn_asymptotic}
\end{align}
where $b_1$, $b_2$ and $c$ are constants, the plus and minus signs on the second equation correspond to the rigid and floppy phases, respectively, and
\begin{equation}
    \tilde{\omega} (\omega)
        \equiv
        \begin{cases}
        & \rho \, \omega^2, \text{ for undamped dynamics,} \\
        & i \, \gamma \, \omega, \text{ for overdamped dynamics,}
        \end{cases}
    \label{eq:OmegaTilde}
\end{equation}
where $\rho$ and $\gamma$ represent the mass density and a drag coefficient, respectively.
Using Eqs.~\eqref{eq:kb_asymptotic} and~\eqref{eq:kn_asymptotic}, we can write,
\begin{equation}
    \kB
        \approx \left[1 + \frac{\DpB}{b\,|\DpRP|\left( \sqrt{1-c \,
            \tilde{\omega} (\omega) / |\DpRP|^2} \pm 1 \right)} \right]^{-1},
    \label{eq:ka_scaling}
\end{equation}
where $b \equiv b_1 \cdot b_2$ is constant. From Eq.~\eqref{eq:ka_scaling}, we can derive the scaling behavior of the frequency-dependent bulk modulus:
\begin{equation}
    B (\omega)
        \approx a \left[1+\frac{\DpB / |\DpRP|^\varphi}{\pazocal{M}_\pm
            (\omega / |\DpRP|^{z \nu}) }\right]^{-1},
    \label{eq:BulkScaling}
\end{equation}
where $a$ is a constant corresponding to the bulk modulus of the fully-populated honeycomb lattice in our model.
The exponent $\varphi=1$, and the product $z\, \nu = 1$ and $2$ for undamped and overdamped dynamics, respectively.
The universal scaling function $\pazocal{M}$ is given by
\begin{equation}
    \pazocal{M}_{\pm} (v)
        \equiv b \times \begin{cases}
        \sqrt{1 - c \, \rho \, v^2} \pm 1, & \text{(undamped),} \\
        \sqrt{1 - i \, c \, \gamma \, v} \pm 1, & \text{(overdamped).}
        \end{cases}
    \label{eq:Mdef}
\end{equation}
where $b$ and $c$ are constants, and the plus and minus correspond to solutions in the elastic and floppy states, respectively.
In Sec.~\ref{subsec:Derivation}, we use Eq.~\eqref{eq:BulkScaling} to derive the scaling behavior of the longitudinal response.

In turn, the scaling behavior for the shear modulus $G$ follows directly from the asymptotic behavior of $\kN$ in Eq.~\eqref{eq:kn_asymptotic}:
\begin{equation}
    G(\omega)
        \approx g \, |\DpRP|^{\beta_\textsc{g}} \pazocal{M}_\pm (\omega / |\DpRP|^{z \nu}),
    \label{eq:ShearScaling}
\end{equation}
where $\beta_\textsc{g}=1$ and $g$ is a constant.

Note that these scaling forms were obtained using an approximation that is valid near the multicritical point J.
Some of the nonuniversal constants change as one moves away from the J point towards larger values of $\DpB$ (see e.g. the slope of the shear modulus in Ref.~\cite{LiarteLub2019}).
Though these constants depend on model details, we expect the functional forms to be universal.

\subsection{Critical exponents and universal scaling function for longitudinal response}
\label{subsec:Derivation}

Now we use the results presented in Sec.~\ref{subsec:Review} to derive the critical
exponents and universal scaling functions for the longitudinal response near both
jamming and RP, for undamped and overdamped dynamics in the solid and fluid phases.

In the long wavelength limit, the longitudinal component of the dynamic response
function $\XL$ of an isotropic viscoelastic material is given
by~\cite{ChaikinLub1995,Bland2016}
\begin{equation}
    \XL
        = \left\{-\rho \, \omega^2 - i \, \gamma \, \omega + q^2 \left[B(\omega)
            + 2 \, \frac{D-1}{D} G(\omega)\right] \right\}^{-1}.
    \label{eq:Response}
\end{equation}
The complex elastic moduli $B(\omega) = B^\prime + i B^{\prime \prime}$ and $G(\omega) = G^\prime + i G^{\prime \prime}$ can be decomposed into storage ($B^\prime$ and $G^\prime$) and loss ($B^{\prime \prime}$ and $G^{\prime \prime}$) components.
Interestingly, we observe a nonzero loss modulus in our effective-medium results even if there is no dissipation term in the dynamics (i.e. if $\gamma = 0$.)
This happens because the dynamical CPA involves an average over disorder that maps simple springs into Kelvin-Voigt elements (combinations of springs and dashpots~\cite{Bland2016}), which lead to nonzero imaginary parts if the frequency is sufficiently high. Physically, this is the way CPA incorporates scattering of long-wavelength phonons off the disordered lattice.

For $G/B \ll 1$ (near jamming) we can write
\begin{align}
    \XL
        & \approx \left[-\rho \, \omega^2 - i \, \gamma \, \omega + q^2 \, B(\omega) \right]^{-1} \nonumber \\
        & \approx \left\{-\rho \, \omega^2 - i \, \gamma \, \omega + a \, q^2 \,
        \right. \nonumber \\
        & \quad \left. \times \left[1 +\frac{\DpB / |\DpRP|^\varphi}{\pazocal{M}_\pm
            (\omega / |\DpRP|^{z \nu}) }\right]^{-1} \right\}^{-1}
    \label{eq:ResponseApp}
\end{align}
where $\varphi=1$, $z \, \nu = 1$ and $2$ for undamped and overdamped dynamics, respectively, and we have used the asymptotic form for the bulk modulus [Eq.~\eqref{eq:BulkScaling}].
Now we multiply both sides of~\eqref{eq:ResponseApp} by $|\DpRP|^2$ to write
\begin{align}
    |\DpRP|^2 \XL
        & \approx \left\{-\rho \left(\frac{\omega}{|\DpRP|}\right)^2 - i \, \gamma \frac{\omega}{|\DpRP|^2}
        \right. \nonumber \\
        & \quad \left. + a \left(\frac{q}{|\DpRP|}\right)^2 \left[1 +\frac{\DpB / |\DpRP|^\varphi}{\pazocal{M}_\pm
            (\omega / |\DpRP|^{z \nu}) }\right]^{-1} \right\}^{-1}.
    \label{eq:ResponseApp2}
\end{align}

Let us define the nonuniversal scaling factors~\cite{Cardy1996},
\begin{align}
    & \chi_0 \equiv c, \quad q_0 \equiv \frac{1}{\sqrt{a c}}, \quad \delta_0 \equiv b, \\
    & \omega_0 \equiv \begin{cases}
        1/\sqrt{\rho \, c}, & \text{for the undamped case,} \\
        1/(\gamma \, c), & \text{for the overdamped case,}
    \end{cases}
\end{align}
which lead to the scaling form:
\begin{equation}
    \frac{\XL}{\chi_0}
        \approx |\DpRP|^{-\gamma} \pazocal{L} \left( \frac{q / q_0}{|\DpRP|^{\nu}},
            \frac{\omega / \omega_0}{|\DpRP|^{z \nu}},
            \frac{\DpB/\delta_0}{|\DpRP|^{\varphi}}\right),
\label{eq:ResponseScaling}
\end{equation}
with
\begin{equation}
    \pazocal{L}( u, v, w )
        = \left[\frac{u^2}{1+w / \left(\displaystyle\sqrt{1-\tilde{v}(v)}\pm 1\right)}
            -\tilde{v} (v)\right]^{-1},
    \label{eq:Xdef}
\end{equation}
where $\pazocal{L}$ is a universal scaling function, the exponents $\gamma=2$ and $\nu=1$ for jamming, and
\begin{equation}
    \tilde{v} (v)
        = \begin{cases}
        v^2, & \text{for the undamped case,} \\
        i \, v, & \text{for the overdamped case.}
        \end{cases}
    \label{eq:uDef}
\end{equation}
The exponents $\gamma$, $z$ and $\phi$ are associated with the susceptibility, correlation time, and crossover behavior, respectively~\cite{Cardy1996,Sethna2006}.
As we show in Sec.~\ref{sec:Quantities}, our exponent $\nu$ for the correlation length is associated with traditional definitions for diverging length scales $\ell^*$ and $\ell_c$ (see e.g.~\cite{BaumgartenVT17}), and should not be confused with exponents for the finite-size scaling of the probability density $\Delta \sim L^{1/\nu}$, as reported in calculations based on the pebble game~\cite{JacobsTho1995}.

At fixed $\DpB$, the limit $\DpRP \rightarrow 0$ leads to RP criticality (cf. Fig.~\ref{fig:PhaseDiagram}).
Thus, we can study the crossover to rigidity percolation by considering the invariant scaling combination $\DpB/|\DpRP|^{\varphi} \gg 1$, so that, from Eq.~\eqref{eq:ResponseApp2}:
\begin{align}
    |\DpRP|^2 \XL
        & \approx \left[-\rho \left(\frac{\omega}{|\DpRP|}\right)^2 - i \, \gamma \frac{\omega}{|\DpRP|^2}
        \right. \nonumber \\
        & \quad \left. + a \left(\frac{q}{|\DpRP|}\right)^2 \frac{|\DpRP|^\varphi}{\DpB }\pazocal{M}_\pm \left(\frac{\omega}{|\DpRP|^{z \nu}}\right) \right]^{-1}.
    \label{eq:ResponseAppRP}
\end{align}
Since $\DpB$ is an irrelevant variable for rigidity percolation, we define $q_0 \equiv \sqrt{\DpB / (a \, c)}$.
Now the term $|\DpRP|^\varphi$ has to be incorporated into the invariant scaling combination for $q$, leading to $\nu=1/2$ for RP.
Since the product $z \, \nu$ depends only on the type of dynamics (undamped or overdamped,) the exponent $z$ must also change for rigidity percolation.
The longitudinal response then behaves as:
\begin{equation}
     \frac{\XL}{\chi_0}
    \approx |\DpRP|^{-\gamma} \bar{\pazocal{L}} \left( \frac{q/q_0}{|\DpRP|^{\nu}},
        \frac{\omega/\omega_0}{|\DpRP|^{z \nu}}\right),
    \label{eq:ResponseScalingRP}
\end{equation}
where
\begin{equation}
    \bar{\pazocal{L}}(u,v)
        = \left[u^2 \left(\displaystyle\sqrt{1-\tilde{v}(v)}\pm 1\right)
            -\tilde{v} (v)\right]^{-1}.
    \label{eq:LbarDef}
\end{equation}

Table~\ref{tab:Exponents} lists the values of the critical exponents $\gamma$, $z$, $\nu$, $\varphi$, $\beta_\textsc{b}$ and $\gamma_\textsc{b}$ for both jamming and RP, and for both undamped and overdamped (between parentheses, if different from undamped) dynamics.
The exponents for the bulk modulus $\beta_\textsc{b}$ and bulk viscosity $\gamma_\textsc{g}$ are defined by equations~\eqref{eq:beta_B} and~\eqref{eq:B_Visc_Exp}, respectively, in Sec.~\ref{sec:Quantities}.
Note that previous studies~\cite{BaumgartenVT17,HexnerNLPoster} of the response of frictionless jammed spheres to a sinusoidal perturbation report exponents $\nu$ that are in-between the ones presented here.
\begin{table}[h!]
\centering
\begin{tabular}{ |l|c|c|c|c|c|c| }
\hline
 & $\gamma$ & $z$ & $\nu$ & $\varphi$ & $\beta_\textsc{b}$ & $\gamma_\textsc{b}$ \\
 \hline
 Jamming & 2 & 1 (2) & 1 & 1 & 0 & 1 (2) \\
 Rigidity Percolation &  2  & 2 (4) & 1/2 & - & 1 & 0 (1) \\
\hline
\end{tabular}
\caption{\label{tab:Exponents} Critical exponents (cf. Eqs.~\eqref{eq:ResponseScaling},~\eqref{eq:ResponseScalingRP},~\eqref{eq:BScaling} and~\eqref{eq:zetaScaling}) extracted from the longitudinal
response function near jamming and rigidity percolation for undamped and overdamped
(between parentheses, if different from undamped) dynamics.}
\end{table}

Our formulation of Eqs.~\eqref{eq:ResponseScaling} and~\eqref{eq:ResponseScalingRP} represents a deliberate effort to emphasize model-independent \emph{(universal)} features.
Note e.g. that our model definition of the non-universal scaling factor $q_0$ is different for jamming and RP; the latter involves a term that increases as one moves away from the jamming multicritical point.
Besides, our formulation allows for the suitable incorporation of analytic corrections to scaling~\cite{AharonyFis1980,AharonyFis1983,Cardy1996,RajuSet2019}, which can be added in a case-by-case basis.
In general, we expect these corrections to appear through the introduction of nonlinear scaling fields,
\begin{align}
    & u_q (q, \omega, \delta_\textsc{j}) = \frac{q}{q_0} + \dots \\
    & u_\omega (q, \omega, \delta_\textsc{j}) = \frac{\omega}{\omega_0} + \dots \\
    & u_\textsc{j} (q, \omega, \delta_\textsc{j}) = \frac{\delta_\textsc{j}}{\delta_{0}} + \dots
\end{align}
which would replace $q/q_0$, $\omega / \omega_0$ and $\delta_\textsc{j} / \delta_0$ in Eqs.~\eqref{eq:ResponseScaling} and~\eqref{eq:ResponseScalingRP}. Here the dots represent higher-order terms and perhaps linear terms in the other variables (rotating the axes). These nonlinear scaling fields can be viewed as the difference between the lab parameters and Nature's natural variables, or as the coordinate transformation removing the (hypothetical) nonlinear terms in the renormalization group to their hyperbolic normal form~\cite{RajuSet2019}.
In order to use our scaling predictions to describe behavior far from the critical point, one must first determine the appropriate scaling fields $u_q$, $u_\omega$ and $u_\textsc{j}$ for the particular system.

\section{Scaling and global asymptotic behavior for longitudinal response}
\label{sec:Global}

Our universal scaling forms for the longitudinal response provide a valuable tool both to investigate invariant critical behavior and to quickly assess the overall global behavior of $\XL$.
In this section, we employ full EMT solutions to validate the universal scaling functions $\pazocal{L}$ and $\bar{\pazocal{L}}$ [see Eqs.~\eqref{eq:Xdef} and~\eqref{eq:LbarDef}], by means of scaling collapse plots for both overdamped (Sec.~\ref{subsec:Overdamped}) and undamped (Sec.~\ref{subsec:Undamped}) dynamics.
We then show how to use our solutions to quickly explore the invariant global behavior exhibited by $\XL$.

\subsection{Overdamped dynamics}
\label{subsec:Overdamped}

\begin{figure}[ht]
\begin{center}
\includegraphics[width=\linewidth]{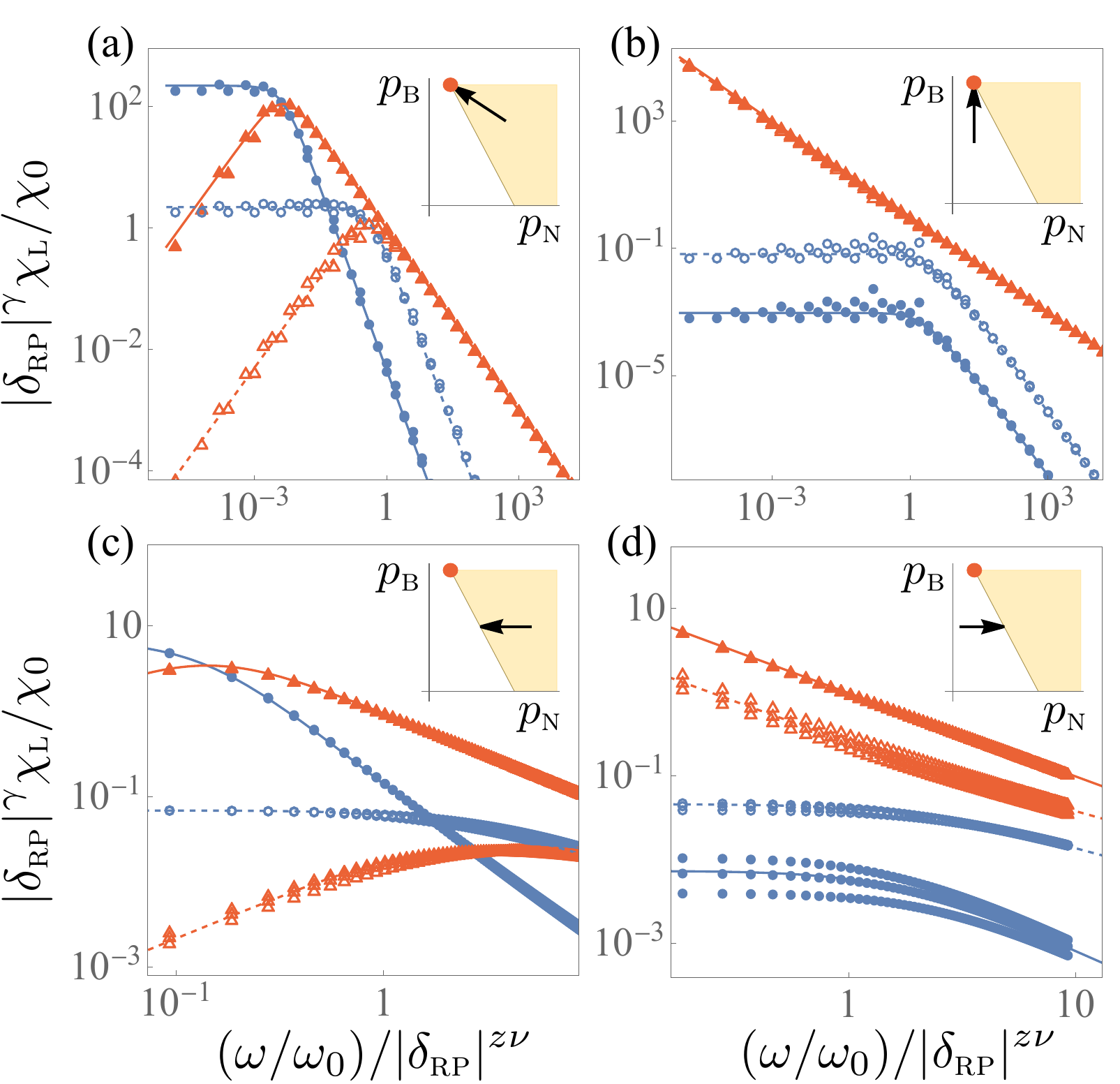} 
\end{center}
\caption{
Scaling collapse plots showing the universal behavior of the longitudinal response as a function of rescaled frequency near jamming (first row) and RP (second row), for \emph{overdamped} dynamics.
Blue disks and red triangles are full solutions of the EMT equations for the real and imaginary parts of $|\DpRP|^\gamma \XL / \chi_0$, respectively.
Solid and dashed curves are the universal scaling predictions of Eqs.~\eqref{eq:Xdef} and~\eqref{eq:LbarDef}.
We consider points approaching jamming and RP along the paths indicated in the inset graphs of each panel.
We use $q/|\DpRP|^{\nu}=0.1$ (closed symbols) and $1$ (open symbols) in all panels, and $\DpB/|\DpRP|^\varphi$ equal to $\sqrt{5}/4$ from the rigid side~(a), and equal to $2$ from the floppy side~(b).
Full solutions run at $|\DpRP| = 10^{-2}$, $10^{-3}$, and $10^{-4}$ for RP and a range $|\DpRP|\in [5\times10^{-2}, 5\times10^{-6}]$ for jamming show convergence to our universal asymptotic predictions.
\label{fig:CollapseOD}}
\end{figure}

Equations~\eqref{eq:ResponseScaling} and~\eqref{eq:ResponseScalingRP} imply that solutions for $|\DpRP|^\gamma \XL$ as a function of one of the three invariant scaling combinations (the other two kept constant) should lie on the curves given by Eqs.~\eqref{eq:Xdef} and~\eqref{eq:LbarDef}, respectively.
Hence, plots for different values of $|\DpRP|$ should collapse for several paths approaching jamming or RP.
Figure~\ref{fig:CollapseOD} shows an example of a scaling collapse plot of the rescaled longitudinal response as a function of rescaled frequency for overdamped dynamics at fixed $q/|\DpRP|^\nu$ and $\delta_\textsc{j}/|\DpRP|^\varphi$, and for paths approaching jamming (first row) and RP (second row) from both the rigid and floppy phases (see inset in each panel).
Although there are model-specific predictions for the nonuniversal scaling factors, we choose them to best fit the collapsed data.

In the elastic phase [(a) and (c)], one observes a crossover to a regime dominated by dissipation (the imaginary part of $\XL$ in red) as the frequency increases.
Note that $\pazocal{L}^\prime$ plateaus at low frequency, but decays to zero at high frequency.
In turn, $\pazocal{L}^{\prime \prime}$ decays to zero both at low and high frequencies, though it decays slower than $\pazocal{L}^\prime$ at large $v$, except in the limit of very large $u$ and $v$, where both $\pazocal{L}^{\prime}$ and $\pazocal{L}^{\prime \prime}$ decay as $v^{-1/2}$.
Thus, there is a frequency scale in which $\pazocal{L}^\prime \sim \pazocal{L}^{\prime \prime}$, characterizing a crossover to a regime where the imaginary dissipative part dominates the dynamic response.
From Eq.~\eqref{eq:Xdef}, we find that $\omega \sim D^* q^2$ in this regime, leading to the definition of an effective diffusion constant
\begin{equation}
    D^*
        \sim |\DpRP|^{(z-2)\nu}.
\end{equation}
Using the exponents shown in Table~\ref{tab:Exponents}, we find that $D^* \sim \pazocal{O}(1)$ and $\sim |\DpRP|$ for jamming and RP, respectively.
In terms of rescaled variables, this crossover happens at $v\sim u^2$ (see Fig. 2(b) and (d) of our companion manuscript~\cite{LiarteSet2021}).
In the liquid phase [(b) and (d)], $\pazocal{L}^\prime$ behaves as in the elastic phase, but $\pazocal{L}^{\prime \prime}$ diverges rather than vanishing at low $v$ due to the predominant viscous response of the fluid state.

\begin{figure*}[!ht]
\begin{center}
\includegraphics[width=\linewidth]{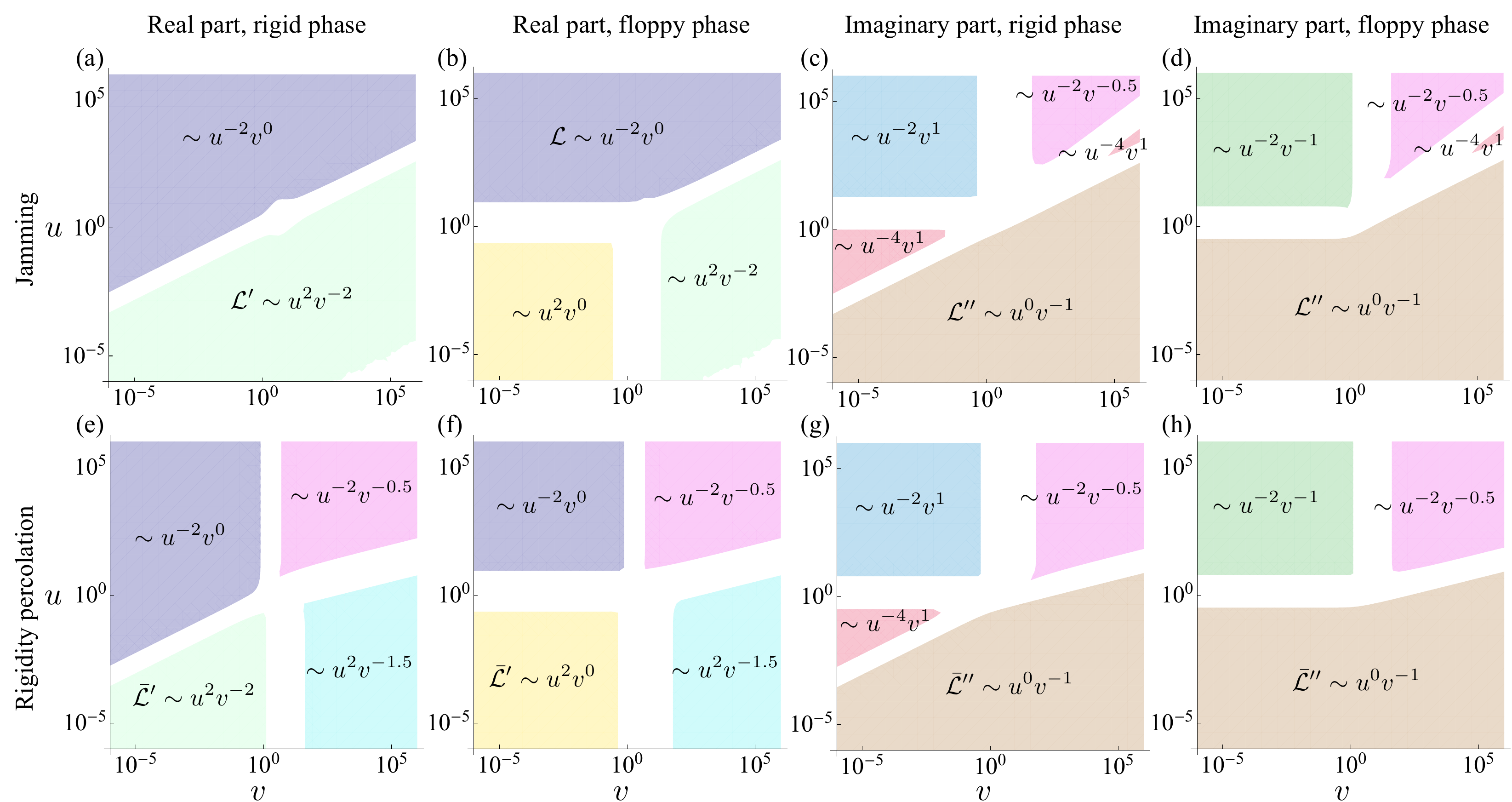}
\end{center}
\caption{{\bf Overdamped asymptotic exponents for universal longitudinal response.} Diagram in the $u$ (rescaled wavevector) $\times v$ (rescaled frequency) plane, showing regions of distinct power-law behavior of the jamming (first row) and RP (second row) universal scaling functions for overdamped dynamics in both the rigid and floppy phases.
The first and second (third and fourth) columns correspond to the real (imaginary) parts of $\pazocal{L}$ and $\bar{\pazocal{L}}$.
We use $w=1$ for jamming.
\label{fig:GlobalPlotsOD}}
\end{figure*}

Equations~(\ref{eq:Xdef}) and (\ref{eq:LbarDef}) also imply that our universal functions for the longitudinal response $\pazocal{L}(u,v,w)$ and $\bar{\pazocal{L}}(u,v)$ generally behave as $u^\alpha v^\beta$ with the exponents $\alpha$ and $\beta$ depending on the region in the $u \textrm{ (rescaled wavector)} \times v \textrm{ (rescaled frequency)}$ plane.
To illustrate and map this global behavior, we show in Fig.~\ref{fig:GlobalPlotsOD} the power-law regions for which $\pazocal{L}(u,v,w) \propto u^\alpha v^\beta$ and $\bar{\pazocal{L}}(u,v) \propto u^\alpha v^\beta$, with $(\alpha, \beta)$ very close to their asymptotic values.
The first and second rows correspond to our scaling forms for jamming and RP, respectively.
To generate each panel, we numerically calculate the exponents using $f_\alpha \equiv \partial \log \pazocal{L}/\partial \log u$ and $f_\beta \equiv \partial \log \pazocal{L}/\partial \log v$ for jamming and similar formulas for RP.
We then plot the regions in which $|f_\alpha - \alpha| < 0.1$ and $|f_\beta - \beta| < 0.1$, for several values of $\alpha$ and $\beta$.

Figure~\ref{fig:GlobalPlotsOD} offers a vivid pictorial view allowing an easier assessment of the global behavior associated with our universal forms for jamming and rigid percolation.
By comparing the two rows, notice how the change in universality class is also reflected in the behavior of the universal scaling functions.
For instance, although jamming and RP exhibit similar qualitative features for the imaginary part [(c), (d), (g) and (h)], RP shows additional regimes for the real part, which do not appear in jamming [compare e.g. (a) and (e) or (b) and (f)].

\subsection{Undamped dynamics}
\label{subsec:Undamped}

\begin{figure}[!ht]
\begin{center}
\includegraphics[height=0.95\linewidth]{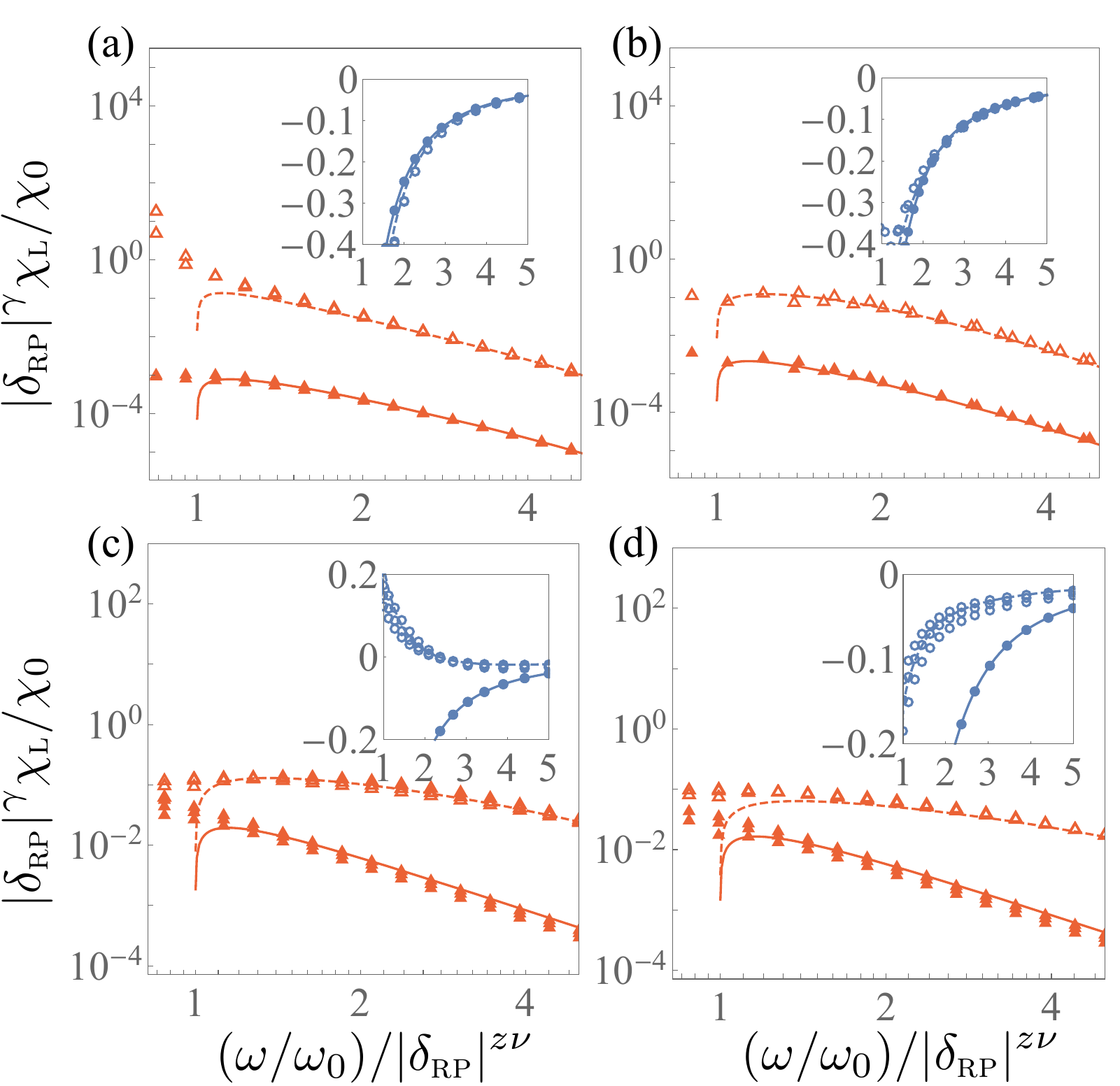}
\end{center}
\caption{Scaling collapse plots showing the universal behavior of the longitudinal response as a function of rescaled frequency near jamming (first row) and RP (second row), for \emph{undamped} dynamics.
Blue disks and red triangles are full solutions of the EMT equations for the real and imaginary parts of $|\DpRP|^\gamma \XL$, respectively.
Solid and dashed curves are the universal scaling predictions of Eqs.~\eqref{eq:Xdef} and~\eqref{eq:LbarDef}.
We consider points approaching jamming and RP along the same paths indicated in the inset graphs of each panel of Fig.~\ref{fig:CollapseOD}; for instance, the path for panel (a) is the same as the one shown in the inset of Fig.~\ref{fig:CollapseOD}(a), etc.
We use $q/|\DpRP|^{\nu}=0.1$ (closed symbols) and $1$ (open symbols) in all panels, and $\DpB/|\DpRP|^\varphi$ equal to $\sqrt{5}/4$ from the rigid side~(a), and equal to $2$ from the floppy side~(b).
Full solutions run at $|\DpRP| = 10^{-2}$, $10^{-3}$, and $10^{-4}$ for both jamming and RP show convergence to our universal asymptotic predictions.
\label{fig:CollapseUD}}
\end{figure}

Now we repeat the analysis of section~\ref{subsec:Overdamped} using $\tilde{\omega} (\omega) = \rho \, \omega^2$ in Eq.~\eqref{eq:OmegaTilde}, i.e. for the undamped case.
Figure~\ref{fig:CollapseUD} shows a scaling collapse plot for the rescaled longitudinal response as a function of rescaled frequency at fixed values of $q/|\DpRP|^\nu$ for jamming and RP (first and second rows, respectively), and at fixed values of $\delta_\textsc{j}/|\DpRP|^\varphi$ for jamming.
We consider several values of $|\DpRP|$ corresponding to points approaching jamming and RP along the paths shown in the insets of Fig.~\ref{fig:CollapseOD}, so that e.g. the path for panel~\ref{fig:CollapseUD}(a) is the same as the one shown in the inset of panel~\ref{fig:CollapseOD}(a).
The real part of $|\DpRP|^\gamma \XL / \chi_0$ can be negative; hence it is shown in linear scale in the insets of each panel of Fig.~\ref{fig:CollapseUD}.
Note that the full solutions of our effective-medium theory equations converge to our universal scaling functions, except in the limit of very low frequencies, which we briefly discuss below.

The asymptotic solutions derived in~\cite{LiarteLub2019} do not capture the small but nonzero imaginary parts of the effective spring constants at frequencies smaller than $\sim \omega^*$ (the characteristic crossover to isostaticity) when there is no damping.
This feature has important consequences for energy dissipation in systems believed to exhibit behavior related to RP.
The corrections to scaling appear as singular perturbations to the self-consistency equations and vanish as powers of $|\DpRP|$ in dimensions larger than three.
Moreover, the scaling variables contain logarithms in two dimensions.
This analysis is beyond the scope of the present work, and will be presented in a separate manuscript.

\begin{figure*}[!ht]
\begin{center}
\includegraphics[width=\linewidth]{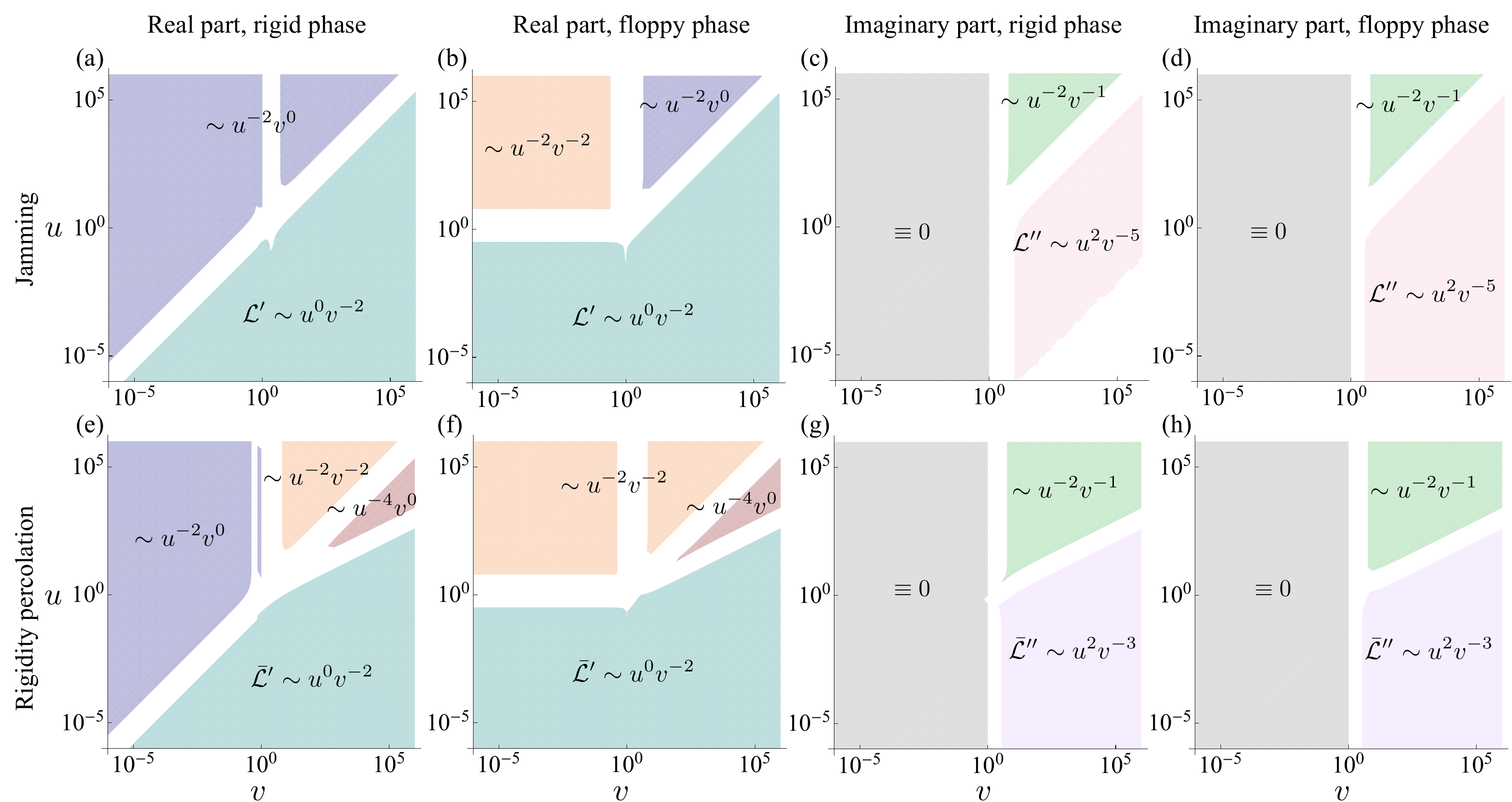}
\end{center}
\caption{{\bf Undamped asymptotic exponents for universal longitudinal response.} Diagram in the $u$ (rescaled wavevector) $\times v$ (rescaled frequency) plane, showing regions of distinct power-law behavior of the jamming (first row) and RP (second row) universal scaling functions for undamped dynamics in both the rigid and floppy phases.
The first and second (third and fourth) columns correspond to the real (imaginary) parts of $\pazocal{L}$ and $\bar{\pazocal{L}}$.
We use the crossover invariant scaling combination $w=1$.
\label{fig:GlobalPlotsUD}}
\end{figure*}

Figure~\ref{fig:GlobalPlotsUD} shows the global asymptotic behavior of our universal scaling functions for the longitudinal response near jamming (first row) and RP (second row).
We use the same approach that we have used to make Fig.~\ref{fig:GlobalPlotsOD}, as described in Sec.~\ref{subsec:Overdamped}.
Notice that the lack of an imaginary part of our scaling functions at low frequency is indicated by gray regions on the left sides of panels (c), (d), (g) and (h).
Again, it is straightforward to compare jamming and RP, or the behavior in the rigid and floppy phases.
As in the overdamped case, the finite bulk modulus at jamming leads to a disparate global behavior of the scaling forms, in comparison with RP.

\section{Derivation of the scaling behavior of other quantities}
\label{sec:Quantities}

Our scaling Ansatz for the longitudinal response [Eqs.~\eqref{eq:ResponseScaling} and~\eqref{eq:ResponseScalingRP}], along with our explicit formulas for the universal functions [Eqs.~\eqref{eq:Xdef} and~\eqref{eq:LbarDef}], allow us to derive more general scaling forms for many quantities.
Here we present a derivation of the universal scaling functions and critical exponents for the transverse dynamic response, bulk and shear moduli, bulk and shear viscosities, density response and correlation functions.

\subsection{Transverse dynamic response}
To extract the universal scaling functions and critical exponents associated with the transverse dynamic response, we follow the same steps that we used in Sec.~\ref{subsec:Derivation}.
We start with the long-wavelength limit of the transverse component of the dynamic response function $\XT$ of an isotropic viscoelastic material~\cite{Bland2016,ChaikinLub1995,Kubo2012,Sethna2006}:
\begin{equation}
    \XT
        = \left\{ -\rho \, \omega^2 - i \gamma \omega + q^2 \, G(\omega) \right\}^{-1}.
    \label{eq:TransverseResponse}
\end{equation}
Near jamming, the complex shear modulus $G(\omega)$ satisfies Eq.~\eqref{eq:ShearScaling}, so that
\begin{align}
    \XT
        & \approx \left\{ -\rho \, \omega^2 - i \gamma \omega + q^2 g |\DpRP|^{\beta_\textsc{g}} \pazocal{M}_{\pm} \left( \frac{\omega}{|\DpRP|^{z \nu}} \right) \right\}^{-1},
\end{align}
where $z\,\nu=1$ and $2$ for undamped and overdamped dynamics, respectively.
Multiplying both sides by $|\DpRP|^2$, we obtain
\begin{align}
    |\DpRP|^2 \XT
        & \approx \left\{ -\rho \, \left(\frac{\omega}{|\DpRP|}\right)^2
            - i \, \gamma \, \frac{\omega}{|\DpRP|^2}
        \right. \nonumber \\
        & \quad + \left. g\,  \left(\frac{q}{|\DpRP|^{1/2}}\right)
            \pazocal{M}_{\pm} \left( \frac{\omega}{|\DpRP|^{z \nu}} \right) \right\}^{-1},
\end{align}
which leads to
\begin{equation}
    \frac{\XT}{\chi_0}
        \approx |\DpRP|^{-\gamma_\textsc{t}} \bar{\pazocal{L}}\left(\frac{q/q_0}{|\DpRP|^{\nu}},
        \frac{\omega/\omega_0}{|\DpRP|^{z \nu}}\right),
    \label{eq:TransverseResponseScaling}
\end{equation}
where $\bar{\pazocal{L}}$ is given by Eq.~\eqref{eq:LbarDef}, and the nonuniversal scaling factors $\chi_0$, $q_0$ and $\omega_0$ \emph{are not} necessarily the same as the ones used for the longitudinal response.
The critical behavior of $\XT$ does not change if one approaches RP instead of jamming.
The exponent $\gamma_\textsc{t}= \gamma = 2$, and the exponents $z$ and $\nu$ are the same as the ones for the longitudinal response near RP (see Table~\ref{tab:Exponents}).

\subsection{Elastic moduli and viscosities}
We have already reviewed the scaling behavior of the elastic moduli in Sec.~\ref{subsec:Review} [see Eqs.~\eqref{eq:BulkScaling} and~\eqref{eq:ShearScaling}].
Here we present an alternative approach that leverages the connection between the dynamic response and the moduli leading to critical exponents and universal scaling functions for $B$ and $G$.

Near jamming, Eq.~\eqref{eq:Response} leads to
\begin{equation}
    B
        \approx \frac{1}{2\,q} \frac{\partial {\XL}^{-1}}{\partial q}.
\end{equation}
Using Eq.~\eqref{eq:ResponseScaling}, we then find,
\begin{align}
    B
        & \approx {\chi_0}^{-1} \frac{1}{2\,q} |\DpRP|^\gamma \frac{\partial {\pazocal{L}}^{-1}}{\partial q}
        \nonumber \\
        & =\frac{(\chi_0 \, {q_0}^2)^{-1}}{2\,((q/q_0)/|\DpRP|^\nu)} |\DpRP|^{\gamma-2\nu} \frac{\partial {\pazocal{L}}^{-1}}{\partial ((q/q_0)/|\DpRP|^\nu)},
\end{align}
so that,
\begin{equation}
    \frac{B}{B_0}
        \approx |\DpRP|^{\beta_\textsc{b}} \pazocal{B} \left( \frac{q/q_0}{|\DpRP|^{\nu}},
            \frac{\omega/\omega_0}{|\DpRP|^{z \nu}},
            \frac{\DpB/\delta_0}{|\DpRP|^{\varphi}}\right),
    \label{eq:BScaling}
\end{equation}
where $B_0$ is a a nonuniversal scaling factor,
\begin{equation}
    \beta_\textsc{b}
        = \gamma - 2\,\nu,
    \label{eq:beta_B}
\end{equation}
and
\begin{equation}
    \pazocal{B}(u, v, w)
        = \frac{1}{2\,u} \frac{\partial}{\partial u} \left[
            \frac{1}{\pazocal{L}(u, v, w)} \right].
    \label{eq:Ydef}
\end{equation}
Near rigidity percolation, the universal function $\pazocal{B}(u,v,w) \rightarrow \bar{\pazocal{B}} (u,v)$, which is given by Eq.\eqref{eq:Ydef} with $\pazocal{L}$ replaced by $\bar{\pazocal{L}}$.

To derive the scaling for the shear modulus, we follow the same steps described in the last paragraph.
Now we explore the connection between $G$ and the transverse response $\XT$.
Equation~\eqref{eq:TransverseResponse} leads to
\begin{equation}
    G
        = \frac{1}{2\,q} \frac{\partial {\XT}^{-1}}{\partial q},
\end{equation}
which is valid near both jamming and RP.
Using Eq.~\eqref{eq:TransverseResponseScaling}, we then find,
\begin{align}
    G
        & \approx \chi_0^{-1} \frac{1}{2\,q} |\DpRP|^{\gamma_\textsc{t}} \frac{\partial {\bar{\pazocal{L}}}^{-1}}{\partial q}
        \nonumber \\
        & = \frac{(\chi_0 {q_0}^2)^{-1}}{2\,((q/q_0)/|\DpRP|^\nu)} |\DpRP|^{\gamma_\textsc{t}-2\nu} \frac{\partial {\bar{\pazocal{L}}}^{-1}}{\partial ((q/q_0)/|\DpRP|^\nu)},
\end{align}
so that,
\begin{equation}
    \frac{G}{G_0}
        \approx |\DpRP|^{\beta_\textsc{g}} \pazocal{G} \left( \frac{q/q_0}{|\DpRP|^{\nu}},
            \frac{\omega/\omega_0}{|\DpRP|^{z \nu}}\right),
    \label{eq:GScaling}
\end{equation}
where $G_0$ is a nonuniversal scaling factor,
\begin{equation}
\beta_\textsc{g} = \gamma_\textsc{t} - 2\,\nu,
\end{equation}
with $\gamma_\textsc{t} = 2$ and $\nu=1/2$ for RP (see Table~\ref{tab:Exponents}).
The universal scaling function
\begin{equation}
    \pazocal{G}(u, v)
        = \frac{1}{2\,u} \frac{\partial}{\partial u} \left[
            \frac{1}{\bar{\pazocal{L}}(u, v)} \right].
    \label{eq:curlyGdef}
\end{equation}
As expected, the scaling of $G$ near both RP and jamming is the same as the scaling behavior of $B$ near RP.

To extract the scaling behavior for the bulk and shear viscosities, we use the definitions $\zeta = B^{\prime \prime} (\omega) / \omega$ and $\eta = G^{\prime \prime} (\omega) / \omega$, so that,
\begin{align}
    \frac{\zeta}{\zeta_0}
        &=|\DpRP|^{-\gamma_\textsc{b}} \pazocal{Z} \left( \frac{q/q_0}{|\DpRP|^{\nu}},
            \frac{\omega/\omega_0}{|\DpRP|^{z \nu}},
            \frac{\DpB/\delta_0}{|\DpRP|^{\varphi}}\right),
    \label{eq:zetaScaling}
    \\
    \frac{\eta}{\eta_0}
        &=|\DpRP|^{-\gamma_\textsc{g}} \pazocal{E} \left( \frac{q/q_0}{|\DpRP|^{\nu}},
            \frac{\omega/\omega_0}{|\DpRP|^{z \nu}}\right),
    \label{eq:etaScaling}
\end{align}
where $\zeta_0$ and $\eta_0$ are nonuniversal scaling factors,
\begin{eqnarray}
    & \gamma_\textsc{b}
        = (2+z) \, \nu - \gamma, \label{eq:B_Visc_Exp}
    \\
    & \gamma_\textsc{g}
        = (2+z) \, \nu - \gamma_\textsc{t},
        \label{eq:gammaG}
\end{eqnarray}
with the exponents $z$ and $\nu$ on the r.h.s. of Eqs.~\eqref{eq:etaScaling} and~\eqref{eq:gammaG} corresponding to the ones listed in Table~\ref{tab:Exponents} for RP, and
\begin{align}
    \pazocal{Z} (u,v,w)
        & = \frac{1}{v} \, \mathrm{Im} \left[\pazocal{B} (u,v,w) \right],
        \label{eq:Zdef} \\
    \pazocal{E} (u,v)
        & = \frac{1}{v} \, \mathrm{Im} \left[\pazocal{G} (u,v) \right],
        \label{eq:Edef}
\end{align}
are the universal scaling functions.
Near RP, $\pazocal{Z} (u,v,w) \rightarrow \bar{\pazocal{Z}} (u,v)$, which is given by Eq.~\eqref{eq:Zdef} with $\pazocal{B}$ replaced by $\bar{\pazocal{B}}$.
As expected, $\pazocal{E}$ does not change near rigidity percolation.

\subsection{Density Response}
The derivation of the density response $\Pi$ proceeds from the equations of motion, in a way that is similar to the derivation of $\XL$~\cite{ChaikinLub1995}.
Whereas $\XL \equiv u_\textsc{l}/f_\textsc{l}$ is defined in Fourier space as the ratio of the longitudinal part of the displacement field $u_\textsc{l}$ to its conjugate external field $f_\textsc{l}$, the density response can be defined as $\Pi \equiv n/h$, where $n$ is the density and $h$ is the density conjugate field.
For small displacements, 
\begin{equation}
    n \equiv n_0\left(1-i \, q \, u_\textsc{l}\right)
    \label{eq:nDef}
\end{equation}
where $n_0$ is a constant given by the average background density.
The appropriate conjugate field in Fourier space that linearly couples to the density in the Hamiltonian is $h \equiv f_\textsc{l}/(i\,q\,n_0)$.
Recasting the equation of motion for $u$ as an equation of motion for $n$ leads to a factor of $q^2$ originating from the divergence operator in Eq.\eqref{eq:nDef} and another factor of $q$ that is present in the definition of $h$.
Thus,
\begin{equation}
    \Pi
        = d^\prime q^2 \XL,
    \label{eq:PiDef0}
\end{equation}
where $d^\prime$ is a constant.

Equations~\eqref{eq:PiDef0} and~\eqref{eq:ResponseScaling} lead to
\begin{equation}
    \Pi
        \approx \chi_0 \,{q_0}^2 d^\prime \left(\frac{q/q_0}{|\DpRP|^\nu} \right)^2
            |\DpRP|^{2\,\nu} \, |\DpRP|^{-\gamma} \pazocal{L},
\end{equation}
resulting in the scaling form:
\begin{equation}
    \frac{\Pi}{\Pi_0}
        \approx |\DpRP|^{-(\gamma-2\nu)} \pazocal{P} \left(
            \frac{q/q_0}{|\DpRP|^{\nu}}, \frac{\omega/\omega_0}{|\DpRP|^{z \nu}},
            \frac{\DpB/\delta_0}{|\DpRP|^{\varphi}}
            \right),
    \label{eq:PiScalingJ}
\end{equation}
which reduces to
\begin{equation}
    \frac{\Pi}{\Pi_0}
        \approx |\DpRP|^{-(\gamma-2\nu)} \bar{\pazocal{P}} \left(
            \frac{q/q_0}{|\DpRP|^{\nu}}, \frac{\omega/\omega_0}{|\DpRP|^{z \nu}} \right),
    \label{eq:PiScalingRP}
\end{equation}
near RP, where $\Pi_0$ is a nonuniversal scaling factor, and
\begin{align}
    \pazocal{P}(u,v,w)
        & = u^2 \pazocal{L}(u,v,w), \\
    \bar{\pazocal{P}} (u,v)
        & = u^2 \bar{\pazocal{L}}(u,v).
    \label{eq:Wdef}
\end{align}

\subsection{Correlation functions}
We end this section with derivations of the scaling behavior of the Ursell function $S_{nn}(q,\omega)$ (the structure factor for isotropic fluids at nonzero $q$) and the scaling behavior of the density-density correlation function in real space: $S_{nn}(r, r^\prime, t, t^\prime) = \langle n(r, t) n(r^\prime, t^\prime)
\rangle -\langle n(r, t) \rangle \langle n(r^\prime, t^\prime) \rangle$.

Using the fluctuation-dissipation theorem~\cite{ChaikinLub1995,Kubo2012},
\begin{equation}
    \omega \, S_{nn}(q,\omega)
        = 2\, T \, \mathrm{Im} \left[\Pi (q,\omega) \right],
\end{equation}
where $T$ is the temperature, and Eq.~\eqref{eq:PiScalingJ}, we obtain
\begin{align}
    S_{nn}(q,\omega)
        & \approx \Pi_0 \frac{2\, T}{\omega} \, \mathrm{Im}
            \left[|\DpRP|^{-(\gamma-2\nu)} \pazocal{P}\right]
    \nonumber \\
        & = \frac{2\, T \, \Pi_0 \, {\omega_0}^{-1}}{(\omega/\omega_0) / |\DpRP|^{z\nu}} \,
            |\DpRP|^{-z\nu-(\gamma-2\nu)} \mathrm{Im}\left[\pazocal{P}\right],
\end{align}
so that
\begin{align}
    \frac{S_{nn} (q, \omega)}{\tilde{S}_0}
        & \approx |\DpRP|^{-[\gamma+(z-2)\nu]}
        \nonumber \\
        & \quad \times \pazocal{S} \left(
            \frac{q/q_0}{|\DpRP|^{\nu}}, \frac{\omega/\omega_0}{|\DpRP|^{z \nu}},
            \frac{\DpB/\delta_0}{|\DpRP|^{\varphi}}
            \right),
    \label{eq:UrsellScalingJ}
\end{align}
which reduces to:
\begin{equation}
    \frac{S_{nn} (q, \omega)}{\tilde{S}_0}
        \approx |\DpRP|^{-[\gamma+(z-2)\nu]} \bar{\pazocal{S}} \left(
            \frac{q/q_0}{|\DpRP|^{\nu}}, \frac{\omega/\omega_0}{|\DpRP|^{z \nu}} \right),
    \label{eq:UrsellScalingRP}
\end{equation}
near RP, where $\tilde{S}_0$ is a nonuniversal scaling factor, and
\begin{align}
    \pazocal{S} (u,v,w)
        & = \frac{1}{v}\mathrm{Im} \left[\pazocal{P}(u, v, w)\right], \\
    \bar{\pazocal{S}} (u,v)
        & = \frac{1}{v}\mathrm{Im} \left[\bar{\pazocal{P}}(u, v)\right].
\end{align}

The two-time density-density correlation function $S_{nn}(r- r^\prime, t- t^\prime)$ is given by,
\begin{align}
    & S_{nn}(r- r^\prime, t- t^\prime)
        = \int d\omega \int d\bm{q} \, e^{-i\omega (t-t^\prime)+i \bm{q}\cdot (r-r^\prime)}
        \nonumber \\
        & \quad \times S_{nn} (q,\omega)
        \nonumber \\
        & = \int d\left(\frac{\omega/\omega_0}{|\DpRP|^{z\nu}}\right)
        \int d\left(\frac{\bm{q}/{q_0}^D}{|\DpRP|^{\nu}}\right) |\DpRP|^{(z + D) \nu}
        \omega_0 \, {q_0}^D
        \nonumber \\
        & \quad \times \exp \left\{i\left[\frac{\bm{q}/q_0}{|\DpRP|^\nu}\cdot \frac{(r-r^\prime)/\ell_0}{|\DpRP|^{-\nu}}-\frac{\omega/\omega_0}{|\DpRP|^{z\nu}} \frac{(t-t^\prime)/t_0}{|\DpRP|^{-z\nu}}\right]\right\}
        \nonumber \\
        & \quad \times S_{nn} (q,\omega),
\end{align}
where $\ell_0\equiv {q_0}^{-1}$ and $t_0 \equiv {\omega_0}^{-1}$ are nonuniversal scaling factors.
Using Eq.~\eqref{eq:UrsellScalingJ}, we then obtain
\begin{align}
    \frac{S_{nn}(r, r^\prime, t, t^\prime)}{S_0}
        & \approx |\DpRP|^{(2+D)\nu-\gamma}
        \nonumber \\
        & \times \pazocal{S}
            \left(\frac{(r-r^\prime)/\ell_0}{|\DpRP|^{-\nu}}, \frac{(t-t^\prime) / t_0}{|\DpRP|^{-z\nu}},
            \frac{\DpB/\delta_0}{|\DpRP|^\varphi}\right),
    \label{eq:C_Scaling}
\end{align}
which reduces to
\begin{align}
    \frac{S_{nn}(r, r^\prime, t, t^\prime)}{S_0}
        & \approx |\DpRP|^{(2+D)\nu-\gamma}
        \nonumber \\
        & \quad \times \bar{\pazocal{S}}
            \left(\frac{(r-r^\prime)/\ell_0}{|\DpRP|^{-\nu}}, \frac{(t-t^\prime)/t_0}{|\DpRP|^{-z\nu}} \right),
    \label{eq:C_Scaling-RP}
\end{align}
near RP, where
\begin{eqnarray}
    & \pazocal{S}(\rho,s,w)
        = \displaystyle\int d\bm{u} \, dv \, e^{i (u \cdot \rho - v s)} \,
            \frac{\mathrm{Im} \, \pazocal{P}(u,v,w)}{v}, \\
    & \bar{\pazocal{S}}(\rho,s)
        = \displaystyle\int d\bm{u} \, dv \, e^{i (u \cdot \rho - v s)} \,
            \frac{\mathrm{Im} \, \bar{\pazocal{P}}(u,v)}{v}.
\end{eqnarray}

Note that Eqs.~\eqref{eq:C_Scaling} and~\eqref{eq:C_Scaling-RP} lead to natural definitions of diverging length and time scales, $\ell = |\DpRP|^{-\nu} \ell_0$ and $\tau = |\DpRP|^{-z\nu} t_0$, respectively.
Interestingly, the time scale divergence is the same for jamming and RP, with $z\nu =1$ for undamped dynamics, and $2$ for overdamped dynamics.
As it should be anticipated, our characteristic length scale diverges as $|\DpRP|^{-1}$ for jamming, and as $|\DpRP|^{-1/2}$ for RP.
These divergences should be compared with traditional definitions of $\ell_c \sim |\Delta z|^{-1/2}$ and $\ell^* \sim |\Delta z|^{-1}$, as discussed in the literature~\cite{EllenbroekSaa2006,LernerWya2014,KarimiMal2015,BaumgartenVT17,HexnerNLPoster}.

\section{Connections with experiments}

The scaling behavior for quantities such as the moduli, viscosities, and correlation functions derived in Sec. \ref{sec:Quantities} provide an approach for experimentally validating our results. These quantities are commonly measured in a number of relevant experimental systems including emulsions, foams, colloidal particles and granular materials \cite{LiuNag2010, mason1995elasticity, saint1999vanishing,zhang2009thermal, majmudar2007jamming, abate2006approach, keys2007measurement}. A careful experiment mapping out how one or more of these quantities evolve upon approaching rigidity could be compared to the critical exponents or even the analytical forms of our universal scaling functions near the jamming and rigidity percolation transitions.

One approach for experimental validation is to directly fabricate disordered elastic networks such as those shown in Fig.~\ref{fig:HTL} using 3D printers or laser-cut 2D sheets \cite{rocks2017designing}. A number of networks could be created at different distances from the RP or jamming point. A set of experiments apply compression or shear could then measure the bulk or shear moduli. Measuring how the moduli change with distance from RP or jamming should then allow experimental measurement of both the critical scaling exponents and the universal functions.

The scaling forms we derive might also apply to common experimental glass-formers such as colloidal suspensions. The density-density correlation function provide an accessible path to experimentally validate our universal scaling function. Recent advances in locating colloidal particles using optical microscopy allow highly precise measurements of particle positions and even local stresses \cite{weeks2000three,lin2016measuring,bierbaum2017light,leahy2018quantitative}. These techniques can be applied to settling system of colloidal silica particles to observe the approach to jamming. Measurements of the two-time density-density correlation function for volume fractions approaching either RP or jamming could be compared to our analytical forms and used to experimentally measure the critical exponents involved. Alternatively, experimental scattering data could allow for relatively easy comparison with our structure factor derivations above. Either approach would allow fitting to our universal scaling functions. 

A number of difficulties still remain for this experimental validation approach. One obstacle is noise in experimental measurements of the density-density correlation function. While state-of-the-art methods for locating particles are very precise, the density-density correlation function still becomes noisy at longer ranges so measurements of correlation scaling will require much care. Additionally, the scaling variables $\DpB$ and $\DpRP$ are defined based on knowledge of the volume fraction for jamming and rigidity percolation. The jamming and rigidity percolation volume fractions are well known for monodisperse hard spheres, but are different for a binary sample (which is often needed to prevent crystalization) and may have to be calculated for any specific sample. Since our universal scaling form applies to a whole class of systems, we remain confident that the right system and experimental protocol could experimentally validate our results.

\section{Summary}
\label{sec:Summary}

In this paper, we have presented a detailed analysis of the universal scaling behavior of disordered viscoelastic materials near the onset of rigidity.
Combining an Ansatz for the longitudinal dynamic response with a semi-analytical effective-medium theory, we have been able to extract critical exponents and explicit formulas for universal scaling functions associated with a variety of quantities, such as elastic moduli, viscosities and correlation functions.
We expect these scaling forms to apply to a large number of systems, from colloidal suspensions and soft gels to the density fluctuations in certain classes of strange metals ~\cite{ThorntonCho2022}.

Possible extensions of our analysis include the incorporation of ingredients such as an \emph{anisotropic distribution of bonds},  which plays an important role in the behavior of colloidal suspensions undergoing shear thickening or shear jamming transitions~\cite{RamaswamyCoh2021}.
Other extensions involve the inclusion of thermal effects~\cite{GiuliWya2015,MaoLub2015}, normal forms or corrections to scaling at the upper critical dimension~\cite{RajuSet2019}, and an annealed~\cite{Cardy1996} or partially-annealed~\cite{CarmoSal2010} distribution of bonds; the latter might be better suited for a description of the fluid phase.
Finally, it would be interesting to investigate the effects on the critical scaling of jamming or RP caused by quenched random fields modeling \emph{active} behavior.

\begin{acknowledgments}
We thank Andrea Liu, Bulbul Chakraborty, Daniel Hexner, Eleni Katifori, Emanuela del Gado, Itay Griniasty, Matthieu Wyart, Meera Ramaswamy, Peter Abbamonte, Sean Ridout, Tom Lubensky and Xiaoming Mao for useful conversations.
This work was supported in part by NSF DMR-1719490 (SJT and JPS), NSF CBET Award \verb|#| 2010118 (DBL, ES, JPS, and IC) and NSF CBET Award \verb|#| 1509308 (ES and IC).
DBL also thanks ICTP-SAIFR for partial financial support through FAPESP grant \verb|#| 2016/01343-7.
DC is supported by a faculty startup grant at Cornell University.
\end{acknowledgments}

\bibliography{References.bib}
\end{document}